\documentclass{llncs}
\usepackage{makeidx}  
\usepackage{url}
\usepackage[cmex10]{amsmath}
\usepackage{amsfonts}
 \usepackage{theorem}
\usepackage[dvips]{epsfig}
\usepackage{enumerate}

 \usepackage{verbatim}
 \usepackage{amssymb}
\usepackage{amsbsy}
 \usepackage{setspace}
\usepackage{url}

\theoremstyle{plain}
\newtheorem{Theorem}{Theorem}

\newtheorem{Proposition}{Proposition}
\newtheorem{Corollary}{Corollary}

\newtheorem{Problem}{Problem}
\newtheorem{Fact}{Fact}

{\theorembodyfont{\rmfamily} \newtheorem{Remark}{Remark}}
{\theorembodyfont{\rmfamily} }
{\theorembodyfont{\rmfamily} \newtheorem{Example}{Example}}
{\theorembodyfont{\rmfamily} }
{\theorembodyfont{\rmfamily} }

\newcommand {\R}{\mathbb R}

\newcommand{\be}{\begin{equation}}
\newcommand{\ee}{\end{equation}}

\newcommand{\Int}{\operatorname{{\mathrm int}}}

\newcommand{\LMD}{\lambda_0,\dots,\lambda_n}

\begin{document}
\mainmatter              

\title{Maximizing  Protein Translation Rate in the Nonhomogeneous Ribosome Flow Model:
 A Convex Optimization
 Approach\thanks{This research is partially supported by research grants from the ISF and from the Ela Kodesz Institute for Medical Engineering and Physical Sciences.}}

\author{Gilad Poker\inst{1} \and Yoram Zarai\inst{2} \and Michael Margaliot\inst{3} \and Tamir Tuller\inst{4} }
\institute{School of EE-Systems, Tel Aviv University, Tel Aviv 69978, Israel \\
\email{pgilad08@gmail.com}
\and
School of EE-Systems, Tel Aviv University, Tel Aviv 69978, Israel \\
\email{yoramzar@mail.tau.ac.il}
\and
School of EE-Systems and the Sagol School of Neuroscience, Tel Aviv University, Tel Aviv 69978, Israel \\
\email{michaelm@post.tau.ac.il}
\and
Dept. of Biomedical Eng. and the Sagol School of Neuroscience, Tel Aviv University, Tel Aviv 69978, Israel\\
\email{tamirtul@post.tau.ac.il}  }

\maketitle

\begin{abstract}
  Translation is an  important stage in gene expression.
  During this stage,  macro-molecules called ribosomes  travel along the mRNA strand
  linking amino-acids together in a specific order to create a functioning  protein.

 An important question, related to many biomedical disciplines, is
 how to maximize
protein production.
  Indeed, translation is known to consume most of the cell's energy and
  it is natural to assume that evolution shaped this process
  so that it maximizes  the protein production rate. If this is indeed so
  then one can estimate various parameters of  the translation machinery
  by solving an appropriate mathematical optimization problem.
  The same problem  also arises in the context of synthetic biology, namely,
  re-engineer
 heterologous genes in order
 to maximize their translation rate in a  host organism.

We consider the  problem of maximizing
the protein production  rate     using a computational model for translation-elongation
called the ribosome flow model~(RFM). This model
describes the flow of the ribosomes along an
 mRNA chain of length~$n$ using a set of~$n$ first-order nonlinear ordinary differential
 equations. It also includes~$n+1$ positive parameters: the ribosomal initiation rate into the mRNA chain, and~$n$ elongation rates along the chain sites.

We show that the steady-state translation
 rate in the~RFM is a \emph{strictly concave} function of its parameters.
 This means that the problem of maximizing the translation rate
 under a suitable
  constraint  always admits a unique solution,
  and that this solution can be determined  using highly-efficient algorithms
  for solving convex optimization problems even for large values of~$n$.
  Furthermore, our analysis shows that the optimal translation rate
   can be computed based only on the optimal initiation rate and the elongation rate of the codons near the beginning of the ORF.
 %
We discuss
 some applications of the theoretical results to
  synthetic biology, molecular evolution, and functional genomics.
\end{abstract}

\section{Introduction}
Gene expression is the process by which
the  information encoded in the
genes is used to synthesize
  proteins.
  The two major steps of gene expression are
    the {transcription} of the genetic information from DNA to messenger RNA~(mRNA) by RNA polymerase,
    and the {translation} of the
     mRNA molecules to proteins. During {gene translation}, the  genetic information is deciphered  into proteins by molecular machines called \emph{ribosomes} that move along the mRNA chain
      in a unidirectional manner from the $5'$ end to the $3'$ end~\cite{Alberts2002}. Each triplet of the mRNA  consecutive nucleotides, called a  \emph{codon}, is decoded by a ribosome into a corresponding amino-acid. The rate in which proteins are produced during the translation step is referred to as the \emph{protein production rate} or \emph{translation rate}.

The translation process occurs in all organisms, in almost all cells, and in almost all conditions.
Thus,  understanding   translation has important implications in many scientific disciplines, including medicine, biotechnology, functional genomics, evolutionary biology, and more.
The amount of biological findings related to translation increases at an exponential rate
and this leads to considerable
interest in
   computational models that can   integrate and analyze these findings~(see, e.g.,
   \cite{Zhang1994,Dana2011,Heinrich1980,MacDonald1968,TullerGB2011,Tuller2007,Chu2012,Shah2013,Deneke2013,Racle2013}).

   A fundamental challenge in

   biotechnology and synthetic biology
   is to control the expression of heterologous genes in a host organism
   in order to synthesize new proteins or to improve certain aspects of the host fitness~\cite{Romanos1992,Moks1987,Binnie1997}.
   Computational models of translation are also important in this context, as they allow one
to simulate and analyze the effect of various manipulations
 of the genomic machinery  on the translation process.

A conventional
computational  model of translation-elongation is the \emph{totally asymmetric simple exclusion process}~(TASEP)~\cite{Shaw2003,TASEP_tutorial_2011}. TASEP is a stochastic
 model that
 describes particles moving along a one-dimensional lattice of sites.
The term \emph{totally asymmetric} is used to indicate unidirectional motion along the chain.
 Each  site can be either empty or occupied by a single particle.
   This captures \emph{interaction} between the  particles,
  as a particle in site~$i$ blocks the movement of
  a particle in site~$i-1$. Hence, the term  \emph{simple exclusion}.
  At each time instant, the sites are scanned
  and provided that a site is occupied by a particle
   and the next
   site is empty, the particle  hops to the next site with some probability.
 The two sides of the chain are connected to  particle reservoirs,
 and particles can hop into the chain (if the first site is empty) and out of
 the chain (if the last site is full).  TASEP is a
 fundamental model in non-equilibrium
 statistical mechanics that has been used to model numerous natural
 and artificial  processes~\cite{TASEP_book}.
 Analysis of TASEP is based on determining the probabilities of
 steady-state configurations using matrix products (see the excellent review
 paper~\cite{solvers_guide}).

The \emph{ribosome flow model}~(RFM)~\cite{reuveni}
 is a \emph{deterministic}  model for translation-elongation that can be obtained
 via  a mean-field approximation of TASEP
    (see, e.g.,~\cite[section 4.9.7]{TASEP_book} and \cite[p. R345]{solvers_guide}).
 The RFM for a chain with~$n$
  sites includes~$n$ first-order, nonlinear ordinary differential equations and~$n+1$ positive parameters:
   the initiation rate $\lambda_0$,
    and elongation rates $\lambda_i$, $i=1,2,\dots,n$, between every two  consecutive
     sites.

  There are indications that in some genes all the  elongation rates along the mRNA chain are
  (approximately) equal~\cite{Ingolia2011}. This may be modeled by assuming
   constant elongation rates in the RFM. This yields the
     \emph{homogeneous ribosome flow model}~(HRFM)~\cite{HRFM_steady_state} that
      includes only two positive parameters: the initiation rate~$\lambda_0$ and the constant elongation rate~$\lambda_c$.

 In a previous study~\cite{HRFM_concave},
we have shown that the steady-state protein translation rate in the HRFM, denoted  $R=R(\lambda_0,\lambda_c)$,
is a concave function of the parameters~$\lambda_0,\lambda_c$.
The proof of this result is based on analyzing the Hessian matrix~$H$ of~$R$ in the HRFM.
Note  that~$H$ has dimensions~$2\times 2$ for all~$n$.
However, the assumption of equal elongation rates
is often too strong. For example, it was shown that factors such as
the adaptation of codons to the tRNA pool~\cite{Dana2014,Kudla2009,Dana2012},
folding of the mRNA~\cite{TullerGB2011,Dana2012},
and local amino acid charge~\cite{TullerGB2011,Dana2012,Charneski2013} affect translation elongation speed.
This induces variations between different elongation rates.
 In these cases, the~HRFM is  not a suitable model,
and one must use the~RFM. The steady-state translation rate in the~RFM
is a function of~$n+1$ parameters, i.e., $R=R(\LMD)$.
In this paper, we show that~$R(\LMD)$, is a \emph{strictly concave}
function of its~$n+1$ positive parameters. Here the Hessian
matrix has dimensions~$(n+1)\times(n+1)$,
and it seems that the approach applied in~\cite{HRFM_concave} cannot be extended to handle the~RFM.
 The proof of our main  result is thus based on an entirely new technique.

\begin{figure}[t]
 \centering
 \scalebox{0.65}{\input{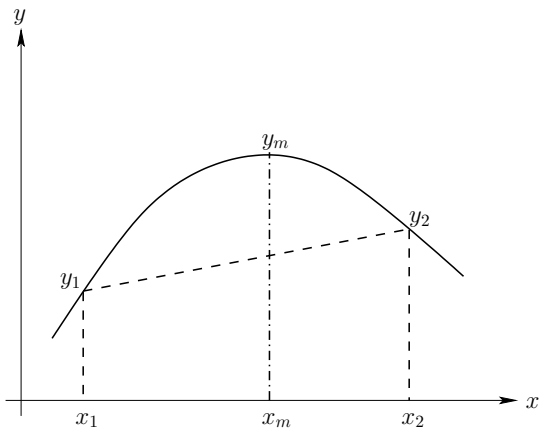}}
\caption{An example of a strictly concave scalar function $y=f(x)$ (solid-line).
 A line segment (dashed-line) between a pair of points ($x_1,y_1$) and ($x_2,y_2$) on the graph lies below the function in the interval between the pair. Note that this function admits a unique maximum point.
 }\label{fig:concave_exp}
\end{figure}

To explain the importance of the strict concavity of~$R$,
consider Fig.~\ref{fig:concave_exp}
that depicts, for simplicity,  a scalar  strictly concave function~$y=f(x)$.
Strict  concavity in this case means the following.
Given any two different values~$x_1$, $x_2$, with corresponding function values $y_1=f(x_1)$ and $y_2=f(x_2)$,
 let~$l=l(x)$ denote the line  that connects the points~$(x_1,y_1)$
 and~$(x_2,y_2)$. Then~$f(x) > l(x)$, for all~$x\in (x_1,x_2)$.
 In other words, the graph of the function lies above the line~$l(x)$.

Concave functions have many useful and desirable properties.
First, a concave function is differentiable almost everywhere.
Second,
 recall that a point~$x_m$ is called
 a \emph{local maximum} of a function~$f$
  if the function values in some  neighborhood of $x_m$
  are smaller than or equal
   to~$f(x_m)$. It is a \emph{global maximum}
    if the function values in its entire domain of definition
  are smaller than or equal
   to~$f(x_m)$.
For a  concave function,
   any \emph{local} maximum is also a \emph{global}
 maximum. If the function is strictly concave then this maximum is unique.

Furthermore,  strict concavity  implies
 that a   simple ``hill climbing'' algorithm can be
 used to find the  global maximum.
 In the depicted one dimensional function, this can be explained  as follows. Select
   an arbitrary point~$x_0$ in the domain of definition
   of~$f$ as a candidate for a maximum point.
Next, determine two points~$x_0^- $ and~$x_0^+$ that are ``close'' to~$x_0$
and satisfy~$x_0^- < x_0 <x_0^+$.
Denote~$y_0=f(x_0)$, $y_0^-=f(x_0^-)$, and~$y_0^+=f(x_0^+)$.
 If $y_0\geq y_0^-$ and~$y_0\geq y_0^+$ then~$x_0$ is a local, and thus global, maximum of the function and the algorithm terminates. Otherwise, at least one of the two values~$y_0^-$, $y_0^+$
 is larger than~$y_0$. The corresponding point, i.e., $x_0^-$ or $x_0^+$,
 becomes the new candidate for a maximum,
  and the algorithm is iterated.
  Under mild assumptions, this simple algorithm is guaranteed to converge
   to the   global maximum of the concave function.
    More generally, there exist highly-efficient algorithms for
     finding the global maximum of multi-dimensional
      concave functions~\cite{convex_boyd}.

A function $g$ is called [strictly]   convex if $ -g$ is [strictly] concave. Thus, the problem of finding the maximum value of a concave function is equivalent to the problem of finding the minimum value of a convex function.
A famous quote by R. T. Rockafellar states
      that:
``...the great watershed in optimization isn't between linearity and nonlinearity, but convexity and nonconvexity.''~\cite{rock_1993}
We note in passing that a linear function $y(x)=ax+b$     is both   concave and   convex.

 Summarizing, our main result implies
 that  the problem of maximizing the protein translation rate, under a simple constraint on the RFM parameter
 values, admits a unique solution, and that this solution can be found numerically using highly-efficient algorithms.
It is important to note that many systems and processes have been modeled and analyzed using TASEP.
These include translation,
  traffic flow, molecular motors, surface growth,
the movement of ants, and more~\cite{TASEP_book}.
All these processes may also be modeled using the RFM, and the problem of maximizing~$R$
seems to be  of importance  in all of them.

We now describe some possible applications
 of the main  result in the context of translation.
A recent work~\cite{Firczuk2013} studied the effect of the intracellular translation factor abundance on the protein production
 rate. Abundance
 of the encoded translation factor was experimentally manipulated to a sub-wild-type level~\cite{Firczuk2013} using the tet07 construct. The reported results suggest that the mapping from levels of translation factors to protein production
  rate is concave (see Fig.~$1$ in \cite{Firczuk2013}).
  This may provide an experimental support to the results presented in this paper. Note that \cite{Firczuk2013} used  the model organism {\em S. cerevisiae} that is
  known to have \emph{non-constant} elongation rates~\cite{TullerGB2011,Dana2014}.
  Thus, the RFM, and not the HRFM~\cite{HRFM_concave},  is a better computational model for describing these experiments.

Translation is known to  consume most of the cell's energy~\cite{Plotkin2011,Tuller2010,Alberts2002}.
 A reasonable assumption is that in
 organisms under strong evolutionary pressure
 the genomic machinery has evolved so that it
  optimizes the translation rate
   given the  available  resources.
   This assumption can be studied in the context of the
   RFM since the concavity of the
    translation rate implies that
    one can easily determine the optimal parameter values,
    and then compare them to  biological findings.
    This may help
      in understanding   the level of selection pressure acting on the genomes
    of various organisms and
    the evolutionary changes in various micro-organisms~\cite{dos-Reis2009}.

In synthetic biology, an important problem is to re-engineer a genetic system by manipulating the transcript sequence, and possibly other intra-cellular variables, in order to obtain an optimal translation rate.
Using our results on the RFM can provide verifiable predictions on how this can be done efficiently.
Another related problem is   optimizing the translation efficiency and protein levels of heterologous genes in a new host~\cite{Plotkin2011,Tuller2010,Gustafsson2004,Kimchi-Sarfaty2013}.
 These genes actually compete with endogenous genes for the available  resources, e.g., initiation factors. Consuming too much resources by the heterologous gene may kill the host~\cite{Plotkin2011,Tuller2010}. Thus, any optimization of the protein translation rate should not consume
  too many resources, as otherwise the fitness of the host may be significantly reduced. This seems to fit well with the constrained optimization problem
 that we pose here  for the RFM.

The remainder of this paper is organized as follows. Section~\ref{sec:rev}
briefly reviews the~RFM. Section~\ref{sec:main} presents the main results.
Section~\ref{sec:discussion}
describes the implications of our results to systems biology, evolution, and synthetic biology,   and describes several possible directions for further research. To streamline the presentation, all the proofs are placed in the Appendix.

\section{Preliminaries}\label{sec:rev}
The RFM~\cite{reuveni} is a  {deterministic} mathematical
model for translation-elongation. In the RFM, mRNA molecules are coarse-grained into a unidirectional chain of $n$ sites of codons.
The RFM is a set of $n$ first-order nonlinear ordinary differential equations:
\begin{align}\label{eq:rfm}
                    \dot{x}_1&=\lambda_{0} (1-x_1) -\lambda_1 x_1(1-x_2), \nonumber \\
                    \dot{x}_2&=\lambda_{1} x_{1} (1-x_{2}) -\lambda_{2} x_{2} (1-x_3) , \nonumber \\
                    \dot{x}_3&=\lambda_{2} x_{ 2} (1-x_{3}) -\lambda_{3} x_{3} (1-x_4) , \nonumber \\
                             &\vdots \nonumber \\
                    \dot{x}_{n-1}&=\lambda_{n-2} x_{n-2} (1-x_{n-1}) -\lambda_{n-1} x_{n-1} (1-x_n), \nonumber \\
                    \dot{x}_n&=\lambda_{n-1}x_{n-1} (1-x_n) -\lambda_n x_n.
\end{align}
Here,~$x_i(t)\in[0,1]$ is the occupancy level at site~$i$ at time~$t$,
normalized so that~$x_i(t)=0$ [$x_i(t)=1$]
implies that site~$i$ is completely empty [completely full] at time~$t$.
The parameter~$\lambda_0 > 0$ is
the initiation rate into the chain,
and~$\lambda_i > 0, i\in\{1,..,n\},$ is a parameter that controls the flow from site~$i$ to site~$i+1$. In particular, $\lambda_n$ controls the output rate at the end of the chain.\footnote{In previous papers on the RFM,
the notation $\lambda$ was used to denote the initiation rate.
Here we use~$\lambda_0$, as this  leads to a more consistent notation.}

\begin{figure}[t]
 \centering
 \scalebox{0.75}{\input{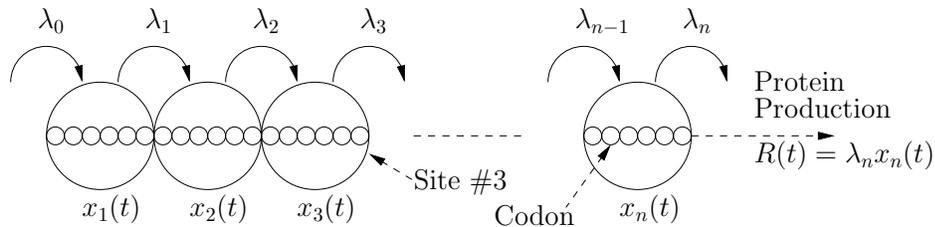}}
\caption{The RFM. Each  site contains a group of codons. The variable~$x_i(t)\in[0,1]$ denotes the normalized
ribosome occupancy level in site $i$ at time $t$. The    initiation rate is denoted~$\lambda_0$  and
 $\lambda_i$  is the transition rate  between   sites~$i$ and~$i+1$. Production rate at time $t$ is $R(t):=\lambda_n x_n(t)$.}\label{fig:rfmm}
\end{figure}

The rate of ribosome flow into
the system is $\lambda_0 (1-x_{1}(t))$.
The rate of ribosome flow   exiting
the last site, i.e., the \emph{protein production rate}, is~$\lambda_{n}  x_{n}(t)$.
The rate of ribosome
flow from site~$i$ to site~$i+1$ is~$\lambda_{i} x_{i}(t)
(1 - x_{i+1}(t) )$  (see Fig.~\ref{fig:rfmm}).
Note that this rate increases with~$x_i(t)$ (i.e., when   site~$i$ is fuller)
and decreases with~$x_{i+1}(t)$ (i.e., when  the consecutive site is becoming fuller).
In this way, the RFM, just like TASEP,
 takes into account the \emph{interaction} between the ribosomes in consecutive
  sites.

  We emphasize that in the RFM the  state-variables take values
  in the closed interval~$[0,1]$ and are not limited to the values~$\{0,1\}$.
   This is different from
  TASEP, where a site can be either empty or full.
  Indeed, the~$x_i$s in the RFM may be interpreted
  as time-averaged occupancy levels in TASEP,
  and this average takes values in~$[0,1]$.

Let~$x(t,a)$ denote the solution of~\eqref{eq:rfm}
at time~$t \ge 0$ for the initial
condition~$x(0)=a$. Since the  state-variables correspond to normalized occupation levels,
  we always assume that~$a$ belongs to the  closed $n$-dimensional
  unit cube:
\[
           C^n:=\{x \in \R^n: x_i \in [0,1] , i=1,\dots,n\}.
\]
It is straightforward to verify that this implies that~$x(t,a) \in C^n$ for all~$t\geq0$.
In other words,~$C^n$ is an invariant set of the dynamics~\cite{RFM_stability}.

Let~$\Int(C^n)$ denote the interior of~$C^n$.
It was shown in~\cite{RFM_stability} that the RFM is a
\emph{monotone dynamical system}~\cite{hlsmith}
and that this implies that~\eqref{eq:rfm}
admits a \emph{unique} equilibrium point~$e \in \Int(C^n)$. Furthermore,
\[
 \lim_{t\to \infty}x(t,a)=e,\quad \text{for all } a \in C^n.
\]
  This means that
all trajectories converge to the steady-state~$e$.

 We note in passing
 that monotone dynamical systems have recently found many
applications in systems biology, see, e.g.,~\cite{angeli_sontag_pnas_2004,mono_chem_2007,sontag_near_2007}
 and the references therein.

For~$x=e$, the left-hand side of all the equations
in~\eqref{eq:rfm} is zero, so
\begin{align} \label{eq:ep}
                      \lambda_0 (1- {e}_1) & = \lambda_1 {e}_1(1- {e}_2)\nonumber \\&
                      = \lambda_2  {e}_2(1- {e}_3)   \nonumber \\ & \vdots \nonumber \\
                    &= \lambda_{n-1} {e}_{n-1} (1- {e}_n) \nonumber \\& =\lambda_n  {e}_n.
\end{align}
Denoting the \emph{steady-state translation rate} by
\be \label{eq:defr}
R:=\lambda_n  {e}_n
\ee
yields
\begin{align}\label{eq:rall}
R=\lambda_i e_i(1-e_{i+1}), \quad i\in\{1,\dots,n\},
\end{align}
where we define~$e_{n+1}:=0$. Also,
\begin{align}\label{eq:list}
                             {e}_n & = R/\lambda_n, \nonumber  \\
                             {e}_{n-1} & = R / (\lambda_{n-1} (1- {e}_n) ),  \nonumber\\
                            & \vdots \nonumber \\
                             {e}_{2} & = R / (\lambda_{2} (1- {e}_3) ), \nonumber\\
                             {e}_{1} & = R / (\lambda_{1} (1- {e}_2) ),
\end{align}
and \be \label{eq:also}
                             {e}_1= 1-R/ \lambda_0.
\ee
Combining~\eqref{eq:list} and~\eqref{eq:also} provides a
\emph{finite continued fraction}~\cite{waad} expression for~$R$:

\begin{align} \label{eq:cf}
                0&= 1-\cfrac{R/ \lambda_0 }
                                  {  1-\cfrac{R / \lambda_1}
                                  {1-\cfrac{R / \lambda_2}{\hphantom{aaaaaaa} \ddots
                             \genfrac{}{}{0pt}{0}{}
                             {1-\cfrac{R/\lambda_{n-1}}{1-R/ \lambda_n.}} }}}
\end{align}
Note that this equation has several solutions for~$R$ (and thus also
several solutions for $e_n=R/\lambda_n)$, however, we are interested only in the unique feasible solution, i.e. the solution corresponding
 to~$e \in \Int(C^n)$.

Eq.~\eqref{eq:cf}
 may be written as~$p(R)=0$, where~$p$ is a  polynomial  of degree~$\lceil (n+1)/2 \rceil$
in~$R$ with coefficients that are algebraic
functions of the~$\lambda_i$s. For example, for~$n=3$,~\eqref{eq:cf} yields
\[
                (\lambda_0 \lambda_2+\lambda_1\lambda_2+\lambda_1\lambda_3)R^2 -(\lambda_0\lambda_1\lambda_2+\lambda_0\lambda_1\lambda_3+\lambda_0\lambda_2\lambda_3+
                \lambda_1\lambda_2\lambda_3)R+\lambda_0\lambda_1\lambda_2\lambda_3=0.
\]

Recent biological findings suggest that in some cases
the transition rate along the mRNA chain is approximately constant \cite{Ingolia2011}.
This may be also the case for gene transcription~\cite{Edri2013}.
To model this, Ref.~\cite{HRFM_steady_state} has considered the RFM in the special case where
\[
            \lambda_1=\lambda_2=\dots=\lambda_n := \lambda_c,
\]
that is, the transition rates~$\lambda_i$, $i=1,2,\dots,n$, are all equal,
and~$\lambda_c$ denotes their common value. Since this \emph{homogeneous ribosome flow model}~(HRFM)
 includes only  two parameters, $\lambda_0$ and~$\lambda_c$,  the analysis is simplified.
In particular,~\eqref{eq:cf} becomes
\begin{align} \label{eq:cfper}
                0&=  1-\cfrac{R/\lambda_0}
                                  {  1-\cfrac{R / \lambda_c}
                                  {1-\cfrac{R / \lambda_c }{\hphantom{aaaaaaa} \ddots
                             \genfrac{}{}{0pt}{0}{}
                             {1-\cfrac{R/\lambda_c}{1-R/ \lambda_c ,}} }}}
\end{align}
where~$\lambda_c$ appears a total of~$n$ times.

Several recent papers analyzed the RFM or HRFM. To model \emph{ribosome recycling}
(see, e.g.,~\cite{recycle2013} and the references therein), Ref.~\cite{RFM_feedback} has considered a closed-loop RFM with a positive linear feedback
from the output~$R$ to the input~$\lambda_0$. It has been shown that the closed-loop
system admits a unique globally asymptotically stable equilibrium point. In~\cite{RFM_entrain}, it has been shown that the state-variables (and thus the protein production
 rate) in the RFM \emph{entrain} to periodically time-varying initiation and/or transition rates. This provides a computational framework for studying entrainment to a periodic excitation (e.g., the cell cycle) at the translation level.
The HRFM with an infinitely-long chain, (i.e. with~$n\to\infty$) was considered in Ref.~\cite{infi_HRFM}. There, a simple closed-form
expression for $e_\infty:=\lim_{n\to \infty}e_n$ was derived, as well as explicit
bounds for $|e_\infty-e_n|$ for all~$n\geq 2$.

In the RFM the steady-state production rate~$R $ is a function of the positive parameters~$\lambda_0,\dots,\lambda_n$.
In this paper, we study the dependence of~$R$ on these parameters.
Our results are based on  a novel, linear-algebraic approach linking the protein translation rate to the maximum eigenvalue of a symmetric, non-negative tridiagonal matrix whose components are  functions of the~$\lambda_i$s.

\section{Main Results}\label{sec:main}
\subsection{Concavity}
 The next result is the main result in this section.
 Recall that all the proofs are placed in the Appendix.
 Let~$\R^{n+1}_+ :=\{ x \in \R^{n+1}:x_i\geq 0, \; i=1,\dots,n+1   \}$.
\begin{Theorem}\label{thm:concave}
Consider the RFM with  dimension~$n$. The steady-state
translation rate $R=R(\LMD)$
is a strictly concave function on $\Int(\R^{n+1}_+)$.
\end{Theorem}

\begin{figure}[t]
  \begin{center}
  \includegraphics[width= 7cm,height=6cm]{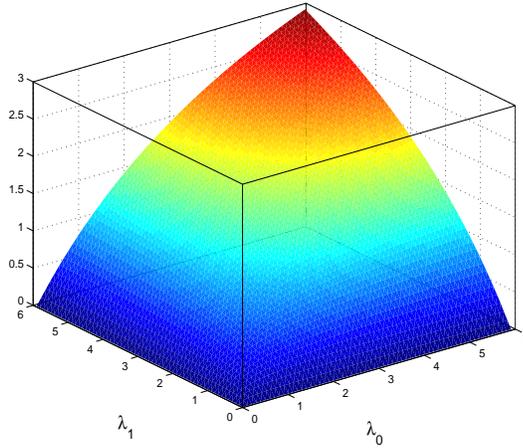}
  \caption{Steady-state translation rate~$R(\lambda_0,\lambda_1)$ in the RFM with
  dimension~$n=1$.  }\label{fig:rfm_n1}
  \end{center}
\end{figure}

The next example demonstrates Theorem~\ref{thm:concave}.
\begin{Example}\label{exa:con2}
Consider the RFM with~$n=1$.
In this case,~\eqref{eq:list} and \eqref{eq:also}
yield~$e_1=R/\lambda_1$ and~$e_1=1-R/\lambda_0$, so
\be\label{eq:rfuncrfm}
                R(\lambda_0,\lambda_1)=
                 \frac{\lambda_0 \lambda_1}{\lambda_0+\lambda_1}.
\ee
Fig.~\ref{fig:rfm_n1} depicts $R(\lambda_0,\lambda_1)$ as a function of its arguments.  It may be seen that this is indeed a strictly concave function on~$\Int(\R^2_+)$.
\end{Example}

Recall that a function~$f:\R^k_+ \to \R$ is called \emph{positively homogeneous of degree~$m$}
if~$f(c x)=c^mf(x)$ for all~$c>0$ and all~$x\in\R^k_+$.
For example, the function~$f(x_1,x_2):=x_1 x_2+ x_2^2$
is positively homogeneous of degree~$2$.
The following result follows immediately from the fact that $R$ always appears in~\eqref{eq:cf} only in terms of the form~$R/\lambda_i$.

\begin{Fact}\label{fact:post_hom}
Consider the RFM with  dimension~$n$.
The function~$R=R(\LMD)$ is positively homogeneous of degree one.
\end{Fact}
In other words,
\be\label{eq:R_is_hom}
            R(c \lambda_0,\dots,c\lambda_n)=c R(\LMD),\quad\text{for all } c>0.
\ee

From a biophysical point of view this
 means that multiplying  the initiation rate
and all  the elongation rates  by the same factor~$c>0$
  increases of the steady-state production  rate by a factor of~$c$.
  This also means that the steady-state occupancy levels~$e_i$, $i=1,2,\dots,n$, remain unchanged with respect to such a  multiplication.

\begin{Example}\label{exa:rfm_n=2}
Consider the RFM with dimension~$n=2$. In this case, the feasible solution of~\eqref{eq:list} and \eqref{eq:also} (i.e., the solution corresponding to a value~$e \in C^2$ for all~$\lambda_0$, $\lambda_1$, $\lambda_2>0$) is
\be\label{eq:rfunc32}
                R(\lambda_0,\lambda_1,\lambda_2)=
                 \frac{\lambda_0\lambda_1+\lambda_0\lambda_2+\lambda_1\lambda_2-\sqrt{(\lambda_0\lambda_1+\lambda_0\lambda_2+\lambda_1\lambda_2)^2-4\lambda_0\lambda_1^2 \lambda_2}}{2\lambda_1},
\ee
and clearly this implies that~$ R(c\lambda_0,c\lambda_1,c\lambda_2)=c R(\lambda_0,\lambda_1,\lambda_2)$.
\end{Example}

Recall that a function~$f:\R^k_+ \to \R$ is called \emph{superadditive}
if~$f(x+y)\geq f(x)+f(y)$
  for all~$x,y\in\R^k_+$.
It is well-known that for a positively homogeneous function,
concavity is equivalent to superadditivity
(see, e.g.,~\cite{fund_convx_ana}). Combining this with Fact~\ref{fact:post_hom} and Theorem~\ref{thm:concave}
 yields the following result.
\begin{Corollary}\label{prop:supadd}
Consider the RFM with dimension~$n$.
The function~$R=R(\LMD)$ is superadditive.
\end{Corollary}
This means that
\[
            R(\lambda_0+\bar \lambda_0,\dots,\lambda_n+\bar \lambda_n) \geq
            R(\LMD )+R(\bar \lambda_0,\dots,\bar \lambda_n),
\]
for all~$\LMD,\bar \lambda_0,\dots,\bar\lambda_n > 0$.
From a biophysical point of view this means the following.
Consider two RFMs,
 one with
 initiation rate~$\lambda_0$ and transition rates~$\lambda_1$, $\lambda_2$, $\dots$, $\lambda_n$,
 and the second with initiation rate~$\bar \lambda_0$ and transition rates~$\bar\lambda_1,\bar\lambda_2,\dots,\bar\lambda_n$.
 The sum of the  production rates of these two RFMs is smaller or equal to the production rate of a single RFM with initiation rate~$\lambda_0+\bar\lambda_0$ and transition rates~$\lambda_1+\bar\lambda_1,\lambda_2+\bar\lambda_2,\dots,\lambda_n+\bar\lambda_n$.
 In other words, a single RFM with rates $\lambda_i+\bar\lambda_i$ is at least as efficient as the total
 of two separate
  RFMs, one with rates $\lambda_i$ and the second with rates $\bar\lambda_i$, for $i=0,1,\dots,n$.

\subsection{Constrained Maximization of the Protein Translation Rate}
Consider the problem of determining the parameter values~$\LMD$ that \emph{maximize}~$R$
(or, equivalently, that \emph{minimize}~$-R$) in the RFM. Obviously, to make this problem meaningful we must constrain the possible parameter values. This leads to the following \emph{constrained optimization problem}.
 \begin{Problem}\label{prob:max}
Given the parameters $w_0,w_1,\dots,w_n, b >0$, minimize~$-R =$ $-R(\LMD)$, with respect to its parameters $\LMD$, subject to the constraints:
\begin{align}\label{eq:constraint}
\sum_{i=0}^n w_i\lambda_i  &\leq b, \\
\LMD &\geq 0.\nonumber
\end{align}
\end{Problem}
 In other words, given an affine constraint on the total rates, namely, the initiation rate $\lambda_0$ and the transition rates $\lambda_1,\dots,\lambda_n$, maximize the protein translation rate. The constraint on $\lambda_i$, $i=0,1,\dots,n$, may be related to factors such as
   the abundance
    of intracellular ribosomes, initiation factors, intracellular tRNA molecules and elongation factors.
The values~$w_i$, $i=0,1,\dots,n$, can be used to provide different weighting  to the different
 rates.

It is not difficult to show that the optimal solution~$\lambda^*$ of
Problem~\ref{prob:max}
always satisfies~$\lambda^* \in \Int(R_+^{n+1})$.
Theorem~\ref{thm:concave} implies that Problem~\ref{prob:max} is a \emph{convex optimization problem}~\cite{convex_boyd}.  It thus enjoys many desirable properties.

The next result shows that increasing any of the $\lambda_i$s increases the translation rate.

\begin{Proposition}\label{prop:pos_derv_rfm}
Consider the RFM with dimension $n$. Then
$\frac{\partial}{\partial \lambda_i}R  >0$ for~$i=0,1,\dots,n$.
\end{Proposition}

In other words,
increasing either the initiation  rate or the elongation rate
at  any site improves the production rate.
\begin{Remark}\label{rem:hom}
Proposition~\ref{prop:pos_derv_rfm} implies that the first constraint  in~\eqref{eq:constraint}
can be replaced by~$\sum_{i=0}^n w_i\lambda_i = b$.
\end{Remark}

\begin{figure}[t]
  \begin{center}
  \includegraphics[width= 8cm,height=7cm]{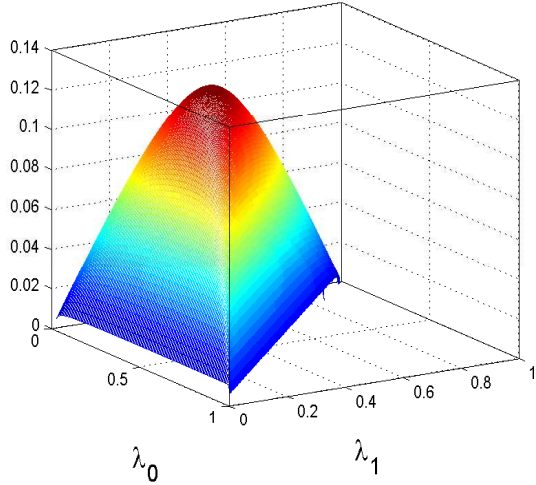}
  \caption{Steady-state translation rate~$R$ in the RFM with dimension~$n=2$ as a function of~$\lambda_0$ and $\lambda_1$ under
  the constraint~$\lambda_2=1-\lambda_0-\lambda_1$. }\label{fig:rfm_n2_c}
  \end{center}
\end{figure}

\begin{Example}\label{exa:rfm2}
Consider Problem~\ref{prob:max} for the RFM with dimension~$n=2$.
In this case,~$R$ is given by~\eqref{eq:rfunc32}.
Let~$b=w_0=w_1= w_2=1$, i.e., the constraint is $\lambda_0+ \lambda_1+\lambda_2  = 1$.
Then $\lambda_2=1-\lambda_0-\lambda_1$, and
substituting this in~\eqref{eq:rfunc32} yields
\begin{align*}
                R =& \frac{\lambda_0\lambda_1+(1- \lambda_0-\lambda_1 )(\lambda_0+\lambda_1)}{2\lambda_1} \\
                &-\frac{\sqrt{(\lambda_0\lambda_1+(1- \lambda_0-\lambda_1 )(\lambda_0+\lambda_1))^2-4\lambda_0\lambda_1^2(1- \lambda_0-\lambda_1 )}}{2\lambda_1}.
\end{align*}
Fig.~\ref{fig:rfm_n2_c} depicts   this function.
It may be seen that~$R=0$ when either~$\lambda_0=0$ or $\lambda_1=0$ (as  a zero  initiation or elongation
rate means of course zero production rate),
and also when~$\lambda_0+\lambda_1=1$ (as then the elongation  rate~$\lambda_2=1-\lambda_0-\lambda_1=0$).
The maximal value,~$R^* = 0.1294$, is obtained
 for~$\lambda_0^* =0.3008$ and $\lambda_1^*  = 0.3984$,
so~$\lambda_2^*=1-\lambda_0^*-\lambda_1^* =0.3008$ (all numbers are to four digit accuracy). Note that~\eqref{eq:rfunc32}
 implies that $R(\lambda_0,\lambda_1,\lambda_2)=R(\lambda_2,\lambda_1,\lambda_0)$ for all $\lambda_0,\lambda_1,\lambda_2>0$, and since the constraint  parameters satisfy $w_0=w_2$, we get $\lambda_0^*=\lambda_2^*$.
\end{Example}

It is clear from~\eqref{eq:rfunc32} that  in general
  an \emph{algebraic} expression for~$R$ in terms of~$\LMD$ does not exist. It
  is possible however to give an algebraic expression for the maximal value~$R^*$ as a function of just
   two optimal parameter values, namely,~$\lambda_0^*$ and $\lambda_1^*$,
    and the parameters in the affine constraint.

\begin{Theorem}\label{thm:mainrfm}
Consider Problem~\ref{prob:max} for the RFM with dimension $n$.
Then
\be \label{eq:mainrfmconst}
R^* = \frac{(\lambda_0^*)^2}{\lambda_0^*+\frac{w_1}{w_0} \lambda_1^*} .
\ee
\end{Theorem}

In other words,
the optimal translation rate~$R^*$    can be computed given the optimal initiation rate
and the first optimal
elongation  rate   (and their corresponding weights
 in the affine constraint). This result holds
 regardless of the length of the  transcript.

It is interesting
  to note that several
  biological studies showed that various signals encoded in the 5'UTR and the \emph{beginning}
   of the ORF can predict the protein levels of endogenous genes with relatively high accuracy~\cite{Kudla2009,Kozak1986,Zur2013,Tuller2010b,Plotkin2011}.

\begin{Example}
Consider again Example~\ref{exa:rfm2}. In this case $ w_1/w_0=1$, $\lambda_0^* = 0.3008$ and $\lambda_1^* =0.3984$, so~\eqref{eq:mainrfmconst} yields
\[
R^* =  \frac{0.3008^2}{0.3008+0.3984}  = 0.1294,
\]
and this agrees with the result in Example~\ref{exa:rfm2}.
\end{Example}

\subsubsection{Maximization with equal constraint weights}
It is interesting to consider the specific
case where all the weights~$w_i$ in the
constrained optimization problem are equal. Indeed, in this case the weights
give equal preference to all the rates, so
if the optimal solution  satisfies~$\lambda_i^* > \lambda_j^*$ for some~$i,j$ then this may be
interpreted as saying
that, in the context of maximizing~$R$,~$\lambda_i$ is ``more important'' than~$\lambda_j$.

\begin{figure}[t]
  \begin{center}
  \includegraphics[width= 8cm,height=7cm]{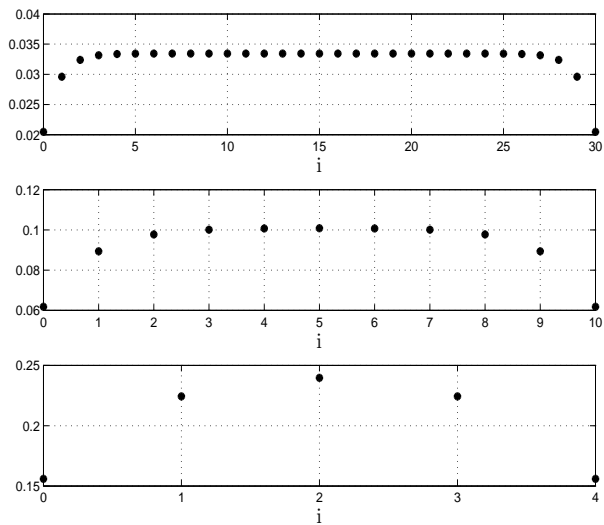}
  \caption{Optimal elongation  rates~$\lambda_i^*$ as a function of~$i$
   for the constraint~$\sum_{i=0}^n \lambda_i=1$.
    Upper figure:~$n=30$; Middle figure:~$n=10$; Lower figure:~$n=4$. }\label{fig:rfm_n_all_l}
  \end{center}
\end{figure}

Fig.~\ref{fig:rfm_n_all_l}
depicts the optimal values~$\lambda_i$ for the case where~$b=1$ and~$w_i=1$ for all~$i$.
In other words, the constraint is~$   \sum_{i=0}^{n} \lambda_i  =  1$. Three cases are shown
corresponding to~$n=30$, $n=10$, and $n=4$.
The optimal values were found numerically using a simple search algorithm
that is guaranteed to converge for convex optimization problems.

 It may be observed that the optimal transition
 rates are symmetric with respect to the index~$i=n/2$.
  In general, the transition  rate~$\lambda_{n/2}^*$
  is larger than all other rates and the optimal values decrease
  as we move towards any edge of the chain.
  The difference between~$\lambda_{n/2}^*$ and~$\lambda_0^*$ (or~$\lambda_n^*$)
  is always visible, but
  the difference between~$\lambda_{n/2}^*$
  and~$\lambda_{(n/2) \pm i}^*$, with~$i$ small,  becomes negligible
   as~$n$ increases.

  Intuitively, these results may be interpreted as follows.
  The importance of an elongation  rate (or the corresponding site)
  depends on its
   ``centrality'', or the mean distance of this site to  other sites in the chain.
     Site~$n/2$  is thus  always the most ``important'' site in the chain. As $n$ increases, the sites near the middle site have almost the same mean distance to the other sites, and   thus become almost as  important.

\begin{figure}[t]
  \begin{center}
  \includegraphics[width= 8cm,height=7cm]{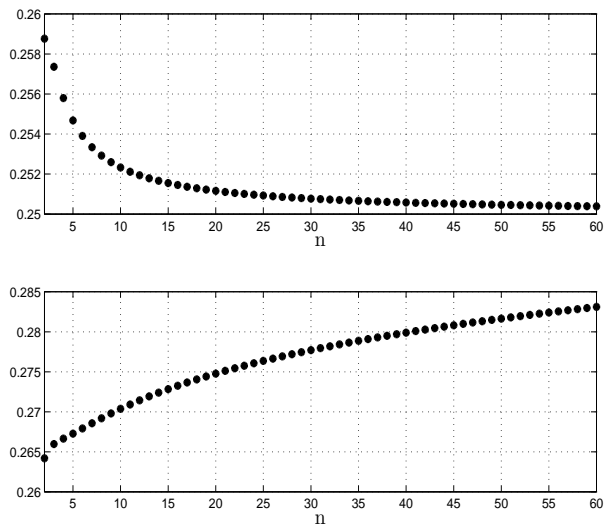}
  \caption{Optimal translation rate $R^*$ as a function of~$n$ for the constraint~$\sum_{i=0}^n \lambda_i=b$.
    Upper figure:~$b=n$; Lower figure:~$b=n^{1.03}$. }\label{fig:rfm_n_bn}
  \end{center}
\end{figure}

Fig.~\ref{fig:rfm_n_bn} depicts the optimal translation rate $R^*$ as a function of $n$ for two different
constraints: $\sum_{i=0}^{n} \lambda_i = n$ and~$\sum_{i=0}^{n} \lambda_i = n^{1.03}$. The first case corresponds to the scenario where the total available resources increases linearly with~$n$ (i.e., $b=n$). It may be observed that in this case the optimal translation rate $R^*$ decreases monotonically with $n$. On the other hand, increasing the total available resources by a rate which is slightly larger than a linear rate (i.e., $b=n^{1.03}$) changes the behavior; $R^*$ in this case increases monotonically with $n$. This result suggests  that in order to maintain the same optimal translation rate value as $n$ increases, the total allocated  resources should increase at a rate that is
 slightly higher than  a linear rate in $n$.

\section{Discussion}\label{sec:discussion}

The RFM is a deterministic mathematical model for translation-elongation.
 It can be derived  via a mean-field
 approximation of a fundamental
  model
 from non-equilibrium statistical  mechanics called
 TASEP. The RFM
  encapsulates both the simple exclusion and the total asymmetry properties of the stochastic TASEP model.
The RFM is characterized by an order $n$, corresponding to the number of sites along the mRNA strand,
a positive initiation rate~$\lambda_0$ and a set of positive alongation rates~$\lambda_1,\dots,\lambda_n$.

In this paper, we show  that the steady-state protein translation rate~$R=R(\LMD)$
in the RFM  is a strictly concave function of its (positive) parameters.
This implies that: (1)~a local maximum of~$R$ is the global maximum (and this maximum is unique); and
(2)~the problem of maximizing the steady-state protein translation rate
under an affine constraint on the RFM parameters is a convex optimization problem.
 Such problems can be solved numerically using highly-efficient algorithms.
The constraint here aims to capture  the limited biosynthetic budget of the cell.

We now describe  the possible  implications of these results
 in  various disciplines including biology,
  synthetic biology, molecular evolution, and functional genomics.
As mentioned above, the functional dependence
of the    translation  rate on
    various variables can also be examined experimentally.
  A recent paper~\cite{Firczuk2013}
  studied the effect of the intracellular translation factor abundance on
protein synthesis.
Experiments based on a {\em tet07} construct were used  to manipulate
the   production of the encoded translation factor to a sub-wild-type level,
  and
  measure
   the translation rate (or protein levels)  for each level of the translation factor(s).
   An analysis of Fig.~$1$ in~\cite{Firczuk2013}
    suggests
 that the mapping from levels of translation factors to
 the translation rate is indeed concave.
Our  results thus provide the first mathematical support for
 the observed concavity in these experiments.

In synthetic biology,
 re-engineering gene expression
 is frequently used to synthesize
  proteins for medical and agricultural goals~\cite{Romanos1992,Moks1987,Binnie1997,Goeddel1979,Mittler2010}.
   For example, in genetically modified crops
   new genes are introduced to the genome of the host
  in order to improve its resistance to certain pests/diseases
   or for improving the nutrient profile of the crop \cite{Mittler2010}.
   Another example is the commercial production of human proteins in
   recombinant microorganisms for therapeutic use~\cite{Romanos1992,Moks1987,Binnie1997,Goeddel1979}.
   This is sometimes based on the natural ability of certain bacteria to efficiently secrete properly folded human proteins (for example, insulin~\cite{Goeddel1979}).
   In this context, a fundamental problem is to
maximize the translation rate of the  heterologous gene
(and thus the protein production rate)
under the given  constraints, e.g., the limited
    availability of   intracellular components involved
  in translation. These constraints are   needed also because
very high initiation and elongation rates     mean that the expression of
 the heterologous gene consumes too much resources
  of the translational machinery (e.g., ribosomes, tRNA molecules, etc),
   thus significantly deteriorating the fitness of the host.
    In addition, very high levels of protein abundance
     may eventually contribute to aggregation of proteins~\cite{Park2008,Kopito2000},
      leading to a decrease in the yield of heterologous protein production.
      All these aspects  are encapsulated in the convex optimization problem that
      is addressed here for the RFM. We believe that   this  mathematical problem
      may thus be used to provide
      verifiable predictions on how to efficiently
      manipulate  the various biological factors.

There is a rich literature on  using optimization theory, combined with
evolutionary arguments, in biology (see, e.g.,~\cite{opt_in_bio_book,may_smith_opt,Alon26092003} and the references therein).
This approach has been often  criticized,
but it has undoubtedly
provided insight into the process of adaption under biological constraints, as well as helped
 to discriminate between alternative hypotheses
for a  suitable ``fitness function''
in various biological mechanisms.
Furthermore,
laboratory evolution experiments showed evolutionary adaptation of
biological processes towards optimal operation levels.
Examples include optimal metabolic fluxes in
{\em E. coli}~\cite{ecoli_optimal}, and
optimal protein expression levels
from the lac operon~\cite{dekel_alon_2005}.
We believe that the optimization problem posed here may lead to further progress in
studying the evolution of the translation machinery.

\begin{figure}[t]
  \begin{center}
  \includegraphics[width= 7cm,height=6cm]{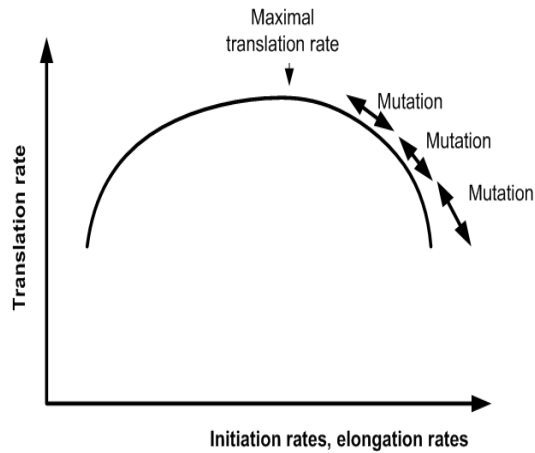}
  \caption{Evolutionary convergence to optimal translational
  state via mutations is similar  to the simple  ``hill climbing'' algorithm. }\label{fig:evolution}
  \end{center}
\end{figure}

The translation machinery is affected by mutations
such as duplication/deletion of a tRNA gene or synonymous mutation affecting the codon bias usage.
The concavity of the translation rate~$R$
may suggest  that the selection of mutations that increase fitness
indeed converges towards the optimal parameter values, as explained by the
simple ``hill climbing'' argument described above (see Fig. \ref{fig:evolution}).

  Recent studies have shown that
   in various organisms the ribosomal density  at the~$5'$ and~$3'$ ends of the ORF
  is higher than
    in the middle of the ORF (see, for example, \cite{Ingolia2009,Ingolia2011,Dana2012,Tuller2010}). In addition, the genomic ribosomal density is relatively constant in the middle of the ORF (usually more than $30$ codons away from the two ORF ends).
The elongation  rate~$\lambda_i$ is negatively
correlated with the ribosomal density (or the probability that a site is occupied)
 at site~$i$. Indeed, if~$\lambda_i$, that controls the
  elongation  rate from  site $i$,
  is small then there  is a higher probability to see a ribosome in this site.
Thus, these biological studies suggest that the elongation  rates at the end of the chain
are lower than in the middle of the chain, and that the rates near the middle
are approximately equal.
This agrees well with the optimal elongation  rates
derived based on our analysis in the case of equal weighting in the constraint (see Fig.~\ref{fig:rfm_n_all_l}).

Our results in the case
 of equal weights in the constraint
 also show that if the total biosynthetic budget~$b(n)$ is a sub-linear or
 linear function of~$n$ then~$R^*$ decreases monotonically with $n$; however, when $b(n)$ grows
  faster than a linear function in $n$ then~$R^*$ \emph{increases} monotonically with $n$.
   The relation between expression levels and gene length has been studied experimentally.
   It has been shown that in some organisms, such as humans and {\em S. cerevisiae} \cite{Eisenberg2003,Dana2011}, expression levels tend to   monotonically decrease with
    gene length (shorter genes have higher expression levels). However, in other organisms, such as plants~\cite{Ren2006}, an opposite relation was reported (longer genes have higher expression levels).
    Our analysis may suggest that one should take into account not only the difference
    in gene length, but also the difference in the available resources  of the translational machinery.

An  interesting question for further research  is whether the translation rate in
 other models  of translation, including various versions of  TASEP  \cite{RFM_feedback,TullerGB2011,Tuller2010,Shah2013,Chu2012,Potapov2012}, is also
 a concave function of its parameters.

Another possible research direction is the design and implementation of biological
experiments based on the analytical results described above. Such experiments
should combine: (1)~methods for manipulating   the translation machinery and/or  the transcript of certain gene(s);
and (2)~online estimation of ribosomal density along the mRNA (e.g. using   ribosome profiling~\cite{ribosome_profiling_rev}).
The elongation
 rate of each codon can be estimated based on a method described in \cite{Dana2014}.
 Manipulation of the translation machinery can include deletion of tRNA genes,
  and   using the tet07 construct to down-regulate the initiation and elongation factors~\cite{Firczuk2013,Bloom-Ackermann2014}.
  Techniques for local manipulation of a transcript include
   generating libraries of a certain heterologous non-functional gene (e.g., a GFP protein).
    In each of the variants a few mutations (relatively to the wild-type) are introduced either in the 5'UTR (corresponding to $\lambda_0$) or the ORF (corresponding to $\lambda_1,\lambda_2,\dots$), and the protein levels and ribosomal densities are measured~\cite{Kudla2009,Welch2009}. The fact that the heterologous gene is non-functional to the host assures that the observed changes in translation efficiency are due to the introduced modifications.

  As a specific example, one  can measure the effect of modifying
   the elongation  rates of different codons  (corresponding to $\lambda_1,\lambda_2,..$) by introducing synonymous mutations in different parts of the ORF. We   expect that a graph depicting the translation rate
  as a function of   elongation rates  will be concave (as in Fig.~\ref{fig:rfm_n2_c}).

Finally, the effect of single mutations in different parts of the transcript on translation rate is a fundamental question related to various biomedical disciplines. Specifically, it is known that codon substitutions in different parts of the coding sequence   affect elongation and initiation rates ({\em i.e.} the $\lambda_i$s) via various mechanisms (e.g. mRNA folding and adaptation to the tRNA pool \cite{dos-Reis2009,Gu2010,Kudla2009,TullerGB2011}). Our result is based on linking~$R$ to the Perron root of a tridiagonal matrix that depends on the~$\lambda_i$s. This can serve as a starting point for sensitivity analysis of~$R$, i.e.
analyzing the effect of small changes in the~$\lambda_i$s on~$R$. This topic is currently under study.

\section*{Appendix: Proofs}
 {\sl Proof of Theorem~\ref{thm:concave}.}
The proof consists of the following steps:
\begin{enumerate}
\item Expressing the term on the right-hand side of~\eqref{eq:cf} as a ratio between two polynomials $p(R)$ and $v(R)$.
\item Linking the numerator polynomial $p(R)$ to the determinant of a symmetric, non-negative tridiagonal matrix $A$ whose entries  depend on the~$\lambda_i$s.
\item Proving that $R^{-1/2}$ is the largest eigenvalue of the matrix $A$.
\item Using the properties of the largest eigenvalue of a symmetric, non-negative
 matrix to show that~$R$ is a strictly concave function of its parameters.
\end{enumerate}

Step 1: Define
\be \label{eq:ijustad23}
g_{n+1}(z,\lambda_0,\dots,\lambda_n):=1-\cfrac{z/ \lambda_0 }
                                  {  1-\cfrac{z / \lambda_1}
                                  {1-\cfrac{z / \lambda_2}{\hphantom{aaaaaaa} \ddots
                             \genfrac{}{}{0pt}{0}{}
                             {1-\cfrac{z/\lambda_{n-1}}{1-z/ \lambda_n }} }}}.
\ee
Then we can rewrite~\eqref{eq:cf} as
\be\label{eq:gn+1R}
g_{n+1}(R,\LMD)=0.
\ee
By the theory of convergents of continued fractions \cite{waad} it follows that
\be\label{eq:gnponq}
g_{n+1}(z,\LMD)=\frac{p_{n+1}(z,\LMD)}{v_{n+1}(z,\lambda_1,\dots,\lambda_n)},
\ee
where $p_{n+1}$ and $v_{n+1}$ are defined recursively by
\begin{align}\label{eq:pndef}
                    p_{-1}(z)&=1,\nonumber \\
                    p_{0}(z)&=1,\nonumber \\
                    p_{k}(z)&=p_{k-1}(z)- z\lambda_{k-1}^{-1} p_{k-2}(z),\quad k \geq 1,
\end{align}
and
\begin{align}\label{eq:qqndef}
                    v_{-1}(z)&=0,\nonumber \\
                    v_{0}(z)&=1,\nonumber \\
                    v_{k}(z)&=v_{k-1}(z)- z\lambda_{k-1}^{-1} v_{k-2}(z),\quad k \geq 1.
\end{align}
For example, for~$n=2$ Eq.~\eqref{eq:ijustad23} yields
\begin{align*}
g_{3}  &= 1-\cfrac{z/ \lambda_0 }
                                  {  1-\cfrac{z / \lambda_1}
                                  {1- {z / \lambda_2} }}\\
                                  &=\frac{ (\lambda_0^{-1} \lambda_2^{-1})z^2 -(\lambda_0^{-1}+\lambda_1^{-1}+\lambda_2^{-1})z  + 1   }
                                  {1-( \lambda_1^{-1}+\lambda_2^{-1} )z } ,
\end{align*}
whereas~\eqref{eq:pndef} and~\eqref{eq:qqndef} yield
\[
            p_3=(\lambda_0^{-1} \lambda_2^{-1})z^2 -(\lambda_0^{-1}+\lambda_1^{-1}+\lambda_2^{-1})z  + 1  ,
\]
and
\[
v_3=1-( \lambda_1^{-1}+\lambda_2^{-1} )z.
\]

Note that~\eqref{eq:pndef} and~\eqref{eq:qqndef} imply that~$p_k=p_k(z,\lambda_0,\dots,\lambda_{k-1})$, $v_k=v_k(z,\lambda_1,\dots,\lambda_{k-1})$,
and
\be\label{eq:pqchar1}
v_{k+1}(z,\lambda_1,\dots,\lambda_k)=p_k(z, \lambda_0,\dots,\lambda_{k-1}) .
\ee

Furthermore, \eqref{eq:also} and~\eqref{eq:list} yield
\be\label{eq:oprr}
       e_k=\frac{p_k(R,\lambda_0,\dots,\lambda_{k-1}) }{ p_{k-1}(R,\lambda_0,\dots,\lambda_{k-2})},\quad k=1,\dots,n,
\ee
so
\begin{align}\label{eq:pkseries}
            p_k &=\frac{p_k } {p_{k-1} }
                     \frac{p_{k-1} }  {p_{k-2} }
                      \dots
                      \frac{p_2 } {p_{1} }  \frac{p_1 } {p_{0} }\nonumber  \\
                     &= e_k e_{k-1}  \dots e_1.
\end{align}
Since $e_i\in(0,1)$ it follows that $p_k(R)\in(0,1)$ for all $k=1,2,\dots,n$.

From~\eqref{eq:gn+1R} and~\eqref{eq:gnponq} it follows that
 \be\label{eq:Lprtt}
\frac{ p_{n+1}(R,\LMD)}{v_{n+1}(R,\lambda_1,\dots,\lambda_n)}  = 0.
\ee
Suppose for a moment that~$v_{n+1}(R,\lambda_1,\dots,\lambda_n)=0$.
Then~\eqref{eq:pqchar1} yields~$ p_n(R, \lambda_0,\dots,\lambda_{n-1})=0$,
and combining this with~\eqref{eq:oprr} yields~$e_n=0$. This is a contradiction, as~$e \in  \Int(C^n)$.
We conclude that the denominator in~\eqref{eq:Lprtt} is not zero, so~\eqref{eq:Lprtt} is well-defined and so
\be\label{eq:enp}
 p_{n+1}(R,\LMD) = 0.
\ee

Step 2: It is well-known that there is a close connection between continued fractions and tridiagonal matrices \cite{wall_contin_frac}. To relate the polynomial~$p_k$ to a tridiagonal matrix, define the polynomials
\be\label{eq:def_poly_qk}
            q_{k } (s): =  s^{k+1} p_k(s^{-2}),\quad k=-1,0,\dots,n+1,
\ee
where $s\in\R \setminus \{0\}$.
Then~\eqref{eq:pndef} yields
\begin{align}\label{eq:queks}
                    q_{0}(s)&=s,\nonumber \\
                    q_{1}(s)&= s^2-\lambda_0^{-1},\nonumber \\
                    q_{k+1}(s)&=s  q_{k}(s)-  \lambda_{k } ^{-1}  q_{k-1}(s),\quad k \geq 1.
\end{align}
Define a $(n+2)\times(n+2)$ Jacobi matrix $A=A(\LMD)$ by
\be\label{eq:bmatrox}
                A:= \begin{bmatrix}
 0 &  \lambda_0^{-1/2}   & 0 &0 & \dots &0&0 \\
\lambda_0^{-1/2} & 0  & \lambda_1^{-1/2}   & 0  & \dots &0&0 \\
 0& \lambda_1^{-1/2} & 0 &  \lambda_2^{-1/2}    & \dots &0&0 \\
 & &&\vdots \\
 0& 0 & 0 & \dots &\lambda_{n-1}^{-1/2}  & 0& \lambda_{n }^{-1/2}     \\
 0& 0 & 0 & \dots &0 & \lambda_{n }^{-1/2}  & 0
 \end{bmatrix}.
\ee
Let~$I_{n+2}$ denote the $(n+2)\times(n+2)$ identity matrix.
Then
\[
sI_{n+2}- A=\begin{bmatrix}
s &  -\lambda_0^{-1/2}   & 0 &0 & \dots &0&0 \\
-\lambda_0^{-1/2} & s  & -\lambda_1^{-1/2}   & 0  & \dots &0&0 \\
 0& -\lambda_1^{-1/2} & s &  -\lambda_2^{-1/2}    & \dots &0&0 \\
 & &&\vdots \\
 0& 0 & 0 & \dots &-\lambda_{n-1}^{-1/2}  & s& -\lambda_{n }^{-1/2}     \\
 0& 0 & 0 & \dots &0 & - \lambda_{n }^{-1/2}  & s
 \end{bmatrix},
\]
and it is straightforward to verify that the determinant of the~$(i+1)\times (i+1)$
leading principal minor of~$sI_{n+2}- A$ is~$q_i(s)$. In particular,~$q_{n+1}(s)=\det(sI_{n+2}-A)$. Combining~\eqref{eq:enp} and~\eqref{eq:def_poly_qk} implies that $q_{n+1}( R^{-1/2}) =0$, so~$R^{-1/2}$ is an eigenvalue of the matrix~$A$.

Step 3: Recall that the \emph{spectral radius} of a square matrix is
   the maximum over the absolute values of its eigenvalue. The spectral radius of a non-negative matrix is an eigenvalue of the matrix called the \emph{Perron root}~\cite{TNM2011}. The next result shows that~$R^{-1/2}$ is the \emph{largest} eigenvalue of the non-negative
   matrix $A$.

\begin{Proposition}\label{prop:perron_root}
           The Perron root of the matrix~$A$ is $R^{-1/2}$.
\end{Proposition}
{\sl Proof of Proposition~\ref{prop:perron_root}.}
It follows from known results on Jacobi matrices (see, e.g.~\cite[Chapter~0]{TNM2011})
that all the eigenvalues of the matrix~$A$ are real and \emph{distinct}, and that if we order them as
\[
                \alpha_1<\alpha_2<\dots<\alpha_{n+2}
\]
then the number of sign changes in the sequence
\[
                               \{ q_n(\alpha_j),\dots,q_0(\alpha_j),1  \}
\]
is~$n+2-j $. Let~$i$ be the index such that~$\alpha_i=R^{-1/2}$. By~\eqref{eq:def_poly_qk},
$q_k(\alpha_i)=R^{-(k+1)/2}p_k( R)$, and~\eqref{eq:pkseries} yields
$q_k(\alpha_i)  >0$ for all $k=0,1,\dots,n$. Thus, the number of sign changes in the sequence
 $\{q_n(\alpha_i),\dots,q_0(\alpha_i),1\}$ is zero, so~$i=n+2$, i.e. $\alpha_{n+2}=R^{-1/2}$.~$\square$

Step 4:   Given a vector~$x=\begin{bmatrix}   x_0& \dots & x_{n }  \end{bmatrix}' \in \Int(\R^{n+1}_+)$,
let~$T(x)$ denote the~$(n+2)\times(n+2)$ tridiagonal matrix whose
main diagonal is zero, and sub- and super-diagonals are the vector~$x$.
Note that this matrix is non-negative and irreducible.
Let~$s_i=s_i(T(x))$, $i=1,\dots,n+2$, denote the  eigenvalues of~$T(x)$ ordered so that
\[
                s_1<s_2<\dots <s_{n+2}.
\]
We already know that $s_{n+2} (T( x^{-1/2} ))=R^{-1/2}( x )$.
Note that
the matrix~$A$ in~\eqref{eq:bmatrox} can be written as~$A=T( \lambda^{-1/2} )$,
where~$ \lambda^{-1/2}:=\begin{bmatrix}   \lambda_{0}^{-1/2}
,\dots,  \lambda_{n }^{-1/2} \end{bmatrix}'$.

Pick~$x,y \in \Int (\R^{n+1}_+)$, with~$x\not = y$, and~$k \in (0,1)$.
Let~$u :=kx +(1-k)y$. Then
\be\label{eq:defrss}
                R(u )=s_{n+2}^{-2} (T( u^{-1/2}   )).
\ee
The function~$f(w):= w^{-1/2}$ is strictly convex on~$w\in (0,\infty)$, so
 \[
                 u^{-1/2}    < k x^{-1/2} +(1-k) y^{-1/2},
 \]
where the inequality between the vectors should be interpreted component-wise.
Since~$T(  u^{-1/2}    )$ is irreducible, this  implies that \cite[Chapter 8]{matrx_ana}
 \[
               s _{n+2} \left  ( T( u^{-1/2}  ) \right ) <   s _{n+2} \left ( Q \right ),
 \]
where~$Q:=T(k x^{-1/2} +(1-k) y^{-1/2})$. Combining this with~\eqref{eq:defrss} yields
 \be\label{eq:nextstep}
              R(u )  >   s _{n+2}^{-2}   ( Q ).
 \ee
 Since $Q$ is  non-negative and symmetric, its induced $2$-norm is
equal to its  spectral radius~$s_{n+2}(Q )$.
 Using the fact that a norm is always convex,
 it is straightforward to see that the norm-squared~$s^2_{n+2}(Q)$ is also convex, so
  \begin{align*}
                 s _{n+2}^{2} \left ( Q \right ) &=
                  s _{n+2}^{2} \left ( kT( x^{-1/2} )+(1-k) T(y^{-1/2}) \right ) \\
                 &\leq
                 k s _{n+2}^{2} \left ( T(  x^{-1/2})   \right ) + (1-k) s _{n+2}^{2} \left (T(   y^{-1/2}) \right ).
 \end{align*}
Combining this with~\eqref{eq:nextstep} yields
\begin{align*}
              R(u ) &> \frac{1}{ k s _{n+2}^{2} \left ( T(  x^{-1/2})   \right ) + (1-k) s _{n+2}^{2} \left (T(   y^{-1/2}) \right ) }\\
                  &\geq  \min\left\{     \frac{1}{   s _{n+2}^{2} \left ( T(  x^{-1/2})   \right )   }   ,    \frac{1}{    s _{n+2}^{2} \left (T(   y^{-1/2}) \right ) }  \right\}.
 \end{align*}
Now~\eqref{eq:defrss}  implies that
  \be\label{eq:stc}
                  R(k x+(1-k)y )  > \min\{     R(x)  ,  R(y)  \},
 \ee
i.e.~$R$ is  strictly quasi-concave on~$\Int(\R^{n+1}_+)$.

Pick~$t\in(0,1)$, and let~$ \mu:=\frac{t R(x)}{ tR(x)+(1-t)R(y) }$. Then~$\mu \in (0,1)$, so~\eqref{eq:stc}
and the homogeneity of~$R$ (see Fact~\ref{fact:post_hom} above) yield
\begin{align*}
                            R\left( \mu \frac{x}{R(x)} + (1-\mu)\frac{y}{R(y)}  \right)&>\min\left\{   R\left( \frac{x}{R(x)}  \right),R\left(\frac{y}{R(y)}\right) \right\}=1.
\end{align*}
Thus,
\[
      R\left ( \frac{ t x+(1-t)y }{  tR(x)+(1-t)R(y) }   \right )                  >1.
\]
Using the homogeneity of~$R$ again gives
\[
      R\left (  t x+(1-t)y      \right )                  >{  tR(x)+(1-t)R(y) },
\]
and this completes the proof of Theorem~\ref{thm:concave}.~$\square$

{\sl Proof of Proposition~\ref{prop:pos_derv_rfm}.}
Let~$v=\begin{bmatrix} v_1 &\dots& v_{n+2} \end{bmatrix}' \in \R^{n+2}_+$ denote a
Perron eigenvector of the symmetric matrix~$A$, i.e., an eigenvector
corresponding to the Perron root~$R^{-1/2}$.
It follows from known results (see, e.g.,~\cite{magnus85}) that
\[
                    \frac{\partial}{\partial\lambda_i} \left(R^{-1/2}\right)= \frac{v'\left(\frac{\partial}{\partial\lambda_i}A\right)v}{v'v},
\]
and combining this with~\eqref{eq:bmatrox} yields
\be\label{eq:derri}
                          \frac{\partial}{\partial\lambda_i} R=  \frac{2R^{ 3/2}  }{ \lambda_i^{3/2} v'v } v_{i+1}v_{i+2} .
\ee
Since all the components of the Perron eigenvector are strictly positive~\cite{matrx_ana}, this implies that~$\frac{\partial}{\partial\lambda_i} R  > 0$ for all~$i = 0 , ... , n$.~$\square$

{\sl Proof of Theorem~\ref{thm:mainrfm}.}
The proof is based on formulating the Lagrangian function associated with Problem~\ref{prob:max} and determining the optimal parameter
values by equating its derivatives to zero (see, e.g.,~\cite{convex_boyd}).
The Lagrangian   is
\[
L(\LMD,\theta):=R(\LMD)+\theta\left(b-\sum_{i=0}^n w_i\lambda_i\right),
\]
where $\theta \in \R$ is the Lagrange multiplier. Differentiating this with respect to $\lambda_i$   and equating to zero yields
\[
\theta w_i=\left( \frac{\partial}{\partial\lambda_i}R \right)  \vert_*,
\]
where $ \vert_*$ means that the equation holds once the optimal
values~$\lambda_i^*$ are substituted.
This implies in particular that
\[
             \frac{w_0}{w_1} =\left(\frac{ \frac{\partial}{\partial\lambda_0}R  } {\frac{\partial}{\partial\lambda_1}R   }  \right) \vert_* ,
\]
and combining this with~\eqref{eq:derri} yields
\be\label{eq:almover}
             \frac{w_0}{w_1} = \frac {(\lambda_1^*)^{3/2} v_1^* } {(\lambda_0^*)^{3/2} v_3^* } ,
\ee
where~$v^*$ is the unique (up to scaling)
Perron eigenvector of the non-negative and irreducible matrix~$A^*:=A(\lambda_0^*,\dots,\lambda_n^*)$ \cite[Ch.~8]{matrx_ana}.
The equation~$A^*v^*=(R^*)^{-1/2} v^*$ yields
\begin{align}\label{eq:Av=Rv}
            (\lambda_0^*)^{-1/2}v_2^*  &=(R^*)^{-1/2} v_1^*, \nonumber \\
            (\lambda_0^*)^{-1/2} v_1^* + (\lambda_1^*)^{-1/2} v_3^* &=(R^*)^{-1/2} v_2^*, \nonumber \\
            (\lambda_1^*)^{-1/2} v_2^* + (\lambda_2^*)^{-1/2} v_4^* &=(R^*)^{-1/2} v_3^*, \nonumber \\
            \vdots \nonumber \\
            (\lambda_n^*)^{-1/2}v_{n+1}^*  &=(R^*)^{-1/2} v_{n+2}^*.
\end{align}
Thus
\[
         (\lambda_1^*) ^{-1/2} v_3^* = (  (R^*)^{-1} (\lambda_0^*)^{1/2}-(\lambda_0^*)^{-1/2}    )v_1^*,
\]
and substituting this in~\eqref{eq:almover} and simplifying yields~\eqref{eq:mainrfmconst}.~$\square$


\bibliographystyle{IEEEtranS}
\bibliography{RFM_bibl_for_rfm_max}

\begin{thebibliography}{10}
\providecommand{\url}[1]{#1}
\csname url@rmstyle\endcsname
\providecommand{\newblock}{\relax}
\providecommand{\bibinfo}[2]{#2}
\providecommand\BIBentrySTDinterwordspacing{\spaceskip=0pt\relax}
\providecommand\BIBentryALTinterwordstretchfactor{4}
\providecommand\BIBentryALTinterwordspacing{\spaceskip=\fontdimen2\font plus
\BIBentryALTinterwordstretchfactor\fontdimen3\font minus
  \fontdimen4\font\relax}
\providecommand\BIBforeignlanguage[2]{{%
\expandafter\ifx\csname l@#1\endcsname\relax
\typeout{** WARNING: IEEEtran.bst: No hyphenation pattern has been}%
\typeout{** loaded for the language `#1'. Using the pattern for}%
\typeout{** the default language instead.}%
\else
\language=\csname l@#1\endcsname
\fi
#2}}

\bibitem{Alberts2002}
B.~Alberts, A.~Johnson, J.~Lewis, M.~Raff, K.~Roberts, and P.~Walter,
  \emph{Molecular Biology of the Cell}.\hskip 1em plus 0.5em minus 0.4em\relax
  New York: Garland Science, 2008.

\bibitem{Alon26092003}
U.~Alon, ``Biological networks: The tinkerer as an engineer,'' \emph{Science},
  vol. 301, no. 5641, pp. 1866--1867, 2003.

\bibitem{angeli_sontag_pnas_2004}
D.~Angeli, J.~E. Ferrell, and E.~D. Sontag, ``Detection of multistability,
  bifurcations, and hysteresis in a large class of biological positive-feedback
  systems,'' \emph{Proceedings of the National Academy of Sciences}, vol. 101,
  pp. 1822--1827, 2004.

\bibitem{fund_convx_ana}
J.~Baptiste, H.~Urruty, and C.~Lemarechal, \emph{Fundamentals of Convex
  Analysis}.\hskip 1em plus 0.5em minus 0.4em\relax Springer, 2001.

\bibitem{Binnie1997}
C.~Binnie, J.~Cossar, and D.~Stewart, ``Heterologous biopharmaceutical protein
  expression in streptomyces,'' \emph{Trends Biotechnol.}, vol.~15, no.~8, pp.
  315--20, 1997.

\bibitem{Bloom-Ackermann2014}
Z.~Bloom-Ackermann, S.~Navon, H.~Gingold, R.~Towers, Y.~Pilpel, and O.~Dahan,
  ``A comprehensive trna deletion library unravels the genetic architecture of
  the trna pool,'' \emph{PLOS Genetics}, vol.~10, no.~1, p. e1004084, 2014.

\bibitem{solvers_guide}
R.~A. Blythe and M.~R. Evans, ``Nonequilibrium steady states of matrix-product
  form: a solver's guide,'' \emph{J. Phys. A: Math. Theor.}, vol.~40, no.~46,
  pp. R333--R441, 2007.

\bibitem{convex_boyd}
S.~Boyd and L.~Vandenberghe, \emph{Convex Optimization}.\hskip 1em plus 0.5em
  minus 0.4em\relax Cambridge University Press, 2004.

\bibitem{Charneski2013}
C.~Charneski and L.~Hurst, ``Positively charged residues are the major
  determinants of ribosomal velocity,'' \emph{PLOS Biology}, vol.~11, no.~3, p.
  e1001508, 2013.

\bibitem{Chu2012}
D.~Chu, N.~Zabet, and T.~von~der Haar, ``A novel and versatile computational
  tool to model translation,'' \emph{Bioinformatics}, vol.~28, no.~2, pp.
  292--3, 2012.

\bibitem{Dana2012}
A.~Dana and T.~Tuller, ``Determinants of translation elongation speed and
  ribosomal profiling biases in mouse embryonic stem cells,'' \emph{PLOS
  Computational Biology}, vol.~8, no.~12, p. e1002755, 2012.

\bibitem{Dana2011}
A.~Dana and T.~Tuller, ``Efficient manipulations of synonymous mutations for
  controlling translation rate--an analytical approach,'' \emph{J. Comput.
  Biol.}, vol.~19, pp. 200--231, 2012.

\bibitem{Dana2014}
A.~Dana and T.~Tuller, ``The effect of {tRNA} levels on decoding times of
  {mRNA} codons,'' \emph{Nucleic Acids Res.}, 2014, to appear.

\bibitem{dekel_alon_2005}
E.~Dekel and U.~Alon, ``Optimality and evolutionary tuning of the expression
  level of a protein,'' \emph{Nature}, vol. 436, pp. 588--592, 2005.

\bibitem{Deneke2013}
C.~Deneke, R.~Lipowsky, and A.~Valleriani, ``Effect of ribosome shielding on
  {mRNA} stability,'' \emph{Phys. Biol.}, vol.~10, no.~4, p. 046008, 2013.

\bibitem{dos-Reis2009}
M.~dos Reis and L.~Wernisch, ``Estimating translational selection in eukaryotic
  genomes,'' \emph{Molecular Biology and Evolution}, vol.~26, no.~2, pp.
  451--61, 2009.

\bibitem{Edri2013}
S.~Edri, E.~Gazit, E.~Cohen, and T.~Tuller, ``The {RNA} polymerase flow model
  of gene transcription,'' \emph{IEEE Trans. Biomed. Circuits Syst.}, vol.~8,
  no.~1, pp. 54--64, 2014.

\bibitem{Eisenberg2003}
E.~Eisenberg and E.~Y. Levanon, ``Human housekeeping genes are compact,''
  \emph{Trends Genet.}, vol.~19, no.~7, pp. 362--5, 2003.

\bibitem{TNM2011}
S.~M. Fallat and C.~R. Johnson, \emph{Totally Nonnegative Matrices}.\hskip 1em
  plus 0.5em minus 0.4em\relax Princeton University Press, 2011.

\bibitem{Firczuk2013}
H.~Firczuk, S.~Kannambath, J.~Pahle, A.~Claydon, R.~Beynon, J.~Duncan,
  H.~Westerhoff, P.~Mendes, and J.~McCarthy, ``An in vivo control map for the
  eukaryotic {mRNA} translation machinery,'' \emph{Mol Syst Biol.}, vol.~9, p.
  635, 2013.

\bibitem{Goeddel1979}
D.~Goeddel, D.~Kleid, F.~Bolivar, H.~Heyneker, D.~Yansura, R.~Crea, T.~Hirose,
  A.~Kraszewski, K.~Itakura, and A.~Riggs, ``Expression in {E}scherichia coli
  of chemically synthesized genes for human insulin,'' \emph{Proceedings of the
  National Academy of Sciences}, vol.~76, no.~1, pp. 106--10, 1979.

\bibitem{Gu2010}
W.~Gu, T.~Zhou, and C.~O. Wilke, ``A universal trend of reduced {mRNA}
  stability near the translation-initiation site in prokaryotes and
  eukaryotes,'' \emph{PLOS Computational Biology}, vol.~6, p. e1000664, 2010.

\bibitem{Gustafsson2004}
C.~Gustafsson, S.~Govindarajan, and J.~Minshull, ``Codon bias and heterologous
  protein expression,'' \emph{Trends Biotechnol.}, vol.~22, pp. 346--353, 2004.

\bibitem{Heinrich1980}
R.~Heinrich and T.~Rapoport, ``Mathematical modelling of translation of {mRNA}
  in eucaryotes; steady state, time-dependent processes and application to
  reticulocytes,'' \emph{J. Theoretical Biology}, vol.~86, pp. 279--313, 1980.

\bibitem{matrx_ana}
R.~A. Horn and C.~R. Johnson, \emph{Matrix Analysis}.\hskip 1em plus 0.5em
  minus 0.4em\relax Cambridge, 2013.

\bibitem{ecoli_optimal}
R.~U. Ibarra, J.~S. Edwards, and B.~O. Palsson, ``Escherichia coli {K-12}
  undergoes adaptive evolution to achieve in silico predicted optimal growth,''
  \emph{Nature}, vol. 420, pp. 186--189, 2002.

\bibitem{ribosome_profiling_rev}
N.~T. Ingolia, ``Ribosome profiling: new views of translation, from single
  codons to genome scale,'' \emph{Nat. Rev. Genet.}, vol.~15, no.~3, pp.
  205--213, 2014.

\bibitem{Ingolia2009}
N.~T. Ingolia, S.~Ghaemmaghami, J.~R. Newman, and J.~S. Weissman, ``Genome-wide
  analysis in vivo of translation with nucleotide resolution using ribosome
  profiling,'' \emph{Science}, vol. 324, no. 5924, pp. 218--23, 2009.

\bibitem{Ingolia2011}
N.~T. Ingolia, L.~Lareau, and J.~Weissman, ``Ribosome profiling of mouse
  embryonic stem cells reveals the complexity and dynamics of mammalian
  proteomes,'' \emph{Cell}, vol. 147, no.~4, pp. 789--802, 2011.

\bibitem{Kimchi-Sarfaty2013}
C.~Kimchi-Sarfaty, T.~Schiller, N.~Hamasaki-Katagiri, M.~Khan, C.~Yanover, and
  Z.~Sauna, ``Building better drugs: developing and regulating engineered
  therapeutic proteins,'' \emph{Trends Pharmacol. Sci.}, vol.~34, no.~10, pp.
  534--548, 2013.

\bibitem{Kopito2000}
R.~Kopito, ``Aggresomes, inclusion bodies and protein aggregation,''
  \emph{Trends Cell Biol.}, vol.~10, no.~12, pp. 524--30, 2000.

\bibitem{Kozak1986}
M.~Kozak, ``Point mutations define a sequence flanking the {AUG} initiator
  codon that modulates translation by eukaryotic ribosomes,'' \emph{Cell},
  vol.~44, no.~2, pp. 283--92, 1986.

\bibitem{Kudla2009}
G.~Kudla, A.~W. Murray, D.~Tollervey, and J.~B. Plotkin, ``Coding-sequence
  determinants of gene expression in {E}scherichia coli,'' \emph{Science}, vol.
  324, pp. 255--258, 2009.

\bibitem{mono_chem_2007}
P.~D. Leenheer, D.~Angeli, and E.~D. Sontag, ``Monotone chemical reaction
  networks,'' \emph{J. Mathematical Chemistry}, vol.~41, pp. 295--314, 2007.

\bibitem{waad}
L.~Lorentzen and H.~Waadeland, \emph{Continued Fractions: Convergence Theory},
  2nd~ed.\hskip 1em plus 0.5em minus 0.4em\relax Paris: Atlantis Press, 2008,
  vol.~1.

\bibitem{MacDonald1968}
C.~T. MacDonald, J.~H. Gibbs, and A.~C. Pipkin, ``Kinetics of biopolymerization
  on nucleic acid templates,'' \emph{Biopolymers}, vol.~6, pp. 1--25, 1968.

\bibitem{magnus85}
J.~R. Magnus, ``On differentiating eigenvalues and eigenvectors,''
  \emph{Econometric Theory}, vol.~1, pp. 179--191, 1985.

\bibitem{RFM_entrain}
M.~Margaliot, E.~D. Sontag, and T.~Tuller, ``Entrainment to periodic initiation
  and transition rates in a computational model for gene translation,''
  \emph{PLoS ONE}, vol.~9, no.~5, p. e96039, 2014.

\bibitem{HRFM_steady_state}
M.~Margaliot and T.~Tuller, ``On the steady-state distribution in the
  homogeneous ribosome flow model,'' \emph{IEEE/ACM Trans. Computational
  Biology and Bioinformatics}, vol.~9, pp. 1724--1736, 2012.

\bibitem{RFM_stability}
M.~Margaliot and T.~Tuller, ``Stability analysis of the ribosome flow model,''
  \emph{IEEE/ACM Trans. Computational Biology and Bioinformatics}, vol.~9, pp.
  1545--1552, 2012.

\bibitem{RFM_feedback}
M.~Margaliot and T.~Tuller, ``Ribosome flow model with positive feedback,''
  \emph{J. Royal Society Interface}, vol.~10, p. 20130267, 2013.

\bibitem{Mittler2010}
R.~Mittler and E.~Blumwald, ``Genetic engineering for modern agriculture:
  challenges and perspectives,'' \emph{Annu Rev Plant Biol.}, vol.~61, pp.
  443--62, 2010.

\bibitem{Moks1987}
T.~Moks, L.~Abrahmsen, E.~Holmgren, M.~Bilich, A.~Olsson, G.~Pohl, C.~Sterky,
  H.~Hultberg, and S.~A. Josephson, ``Expression of human insulin-like growth
  factor {I} in bacteria: use of optimized gene fusion vectors to facilitate
  protein purification,'' \emph{Biochemistry}, vol.~26, no.~17, pp. 5239--44,
  1987.

\bibitem{recycle2013}
E.~Nurenberg and R.~Tampe, ``Tying up loose ends: ribosome recycling in
  eukaryotes and archaea,'' \emph{Trends Biochem Sci.}, vol.~38, no.~2, pp.
  64--74, 2013.

\bibitem{Park2008}
J.~Park, K.~Han, J.~Lee, J.~Song, K.~Ahn, H.~Seo, S.~Sim, S.~Kim, and J.~Lee,
  ``Solubility enhancement of aggregation-prone heterologous proteins by fusion
  expression using stress-responsive escherichia coli protein, {RpoS},''
  \emph{BMC Biotechnol.}, vol.~8, p.~15, 2008.

\bibitem{may_smith_opt}
G.~A. Parker and J.~{Maynard Smith}, ``Optimality theory in evolutionary
  biology,'' \emph{Nature}, vol. 348, pp. 27--33, 1990.

\bibitem{Plotkin2011}
J.~Plotkin and G.~Kudla, ``Synonymous but not the same: the causes and
  consequences of codon bias,'' \emph{Nat. Rev. Genet.}, vol.~12, pp. 32--42,
  2011.

\bibitem{Potapov2012}
I.~Potapov, J.~Makela, O.~Yli-Harja, and A.~S. Ribeiro, ``Effects of codon
  sequence on the dynamics of genetic networks,'' \emph{J. Theoretical
  Biology}, vol. 315, pp. 17--25, 2012.

\bibitem{Racle2013}
J.~Racle, F.~Picard, L.~Girbal, M.~Cocaign-Bousquet, and V.~Hatzimanikatis, ``A
  genome-scale integration and analysis of {Lactococcus} lactis translation
  data,'' \emph{PLOS Computational Biology}, vol.~9, no.~10, p. e1003240, 2013.

\bibitem{Ren2006}
X.~Ren, O.~Vorst, M.~Fiers, W.~Stiekema, and J.~Nap, ``In plants, highly
  expressed genes are the least compact,'' \emph{Trends Genet.}, vol.~22,
  no.~10, pp. 528--32, 2006.

\bibitem{reuveni}
S.~Reuveni, I.~Meilijson, M.~Kupiec, E.~Ruppin, and T.~Tuller, ``Genome-scale
  analysis of translation elongation with a ribosome flow model,'' \emph{PLOS
  Computational Biology}, vol.~7, p. e1002127, 2011.

\bibitem{rock_1993}
R.~T. Rockafellar, ``Lagrange multipliers and optimality,'' \emph{SIAM Review},
  vol.~35, no.~2, pp. 183--238, 1993.

\bibitem{Romanos1992}
M.~Romanos, C.~Scorer, and J.~Clare, ``Foreign gene expression in yeast: a
  review,'' \emph{Yeast}, vol.~8, no.~6, pp. 423--88, 1992.

\bibitem{opt_in_bio_book}
R.~Rosen, \emph{Optimality Principles in Biology}.\hskip 1em plus 0.5em minus
  0.4em\relax London: Butterworths, 1967.

\bibitem{TASEP_book}
A.~Schadschneider, D.~Chowdhury, and K.~Nishinari, \emph{Stochastic Transport
  in Complex Systems: From Molecules to Vehicles}.\hskip 1em plus 0.5em minus
  0.4em\relax Elsevier, 2011.

\bibitem{Shah2013}
P.~Shah, Y.~Ding, M.~Niemczyk, G.~Kudla, and J.~Plotkin, ``Rate-limiting steps
  in yeast protein translation,'' \emph{Cell}, vol. 153, no.~7, pp. 1589--601,
  2013.

\bibitem{Shaw2003}
L.~B. Shaw, R.~K. Zia, and K.~H. Lee, ``Totally asymmetric exclusion process
  with extended objects: a model for protein synthesis,'' \emph{Phys. Rev. E
  Stat. Nonlin. Soft. Matter Phys.}, vol.~68, p. 021910, 2003.

\bibitem{hlsmith}
H.~L. Smith, \emph{Monotone Dynamical Systems: An Introduction to the Theory of
  Competitive and Cooperative Systems}, ser. Mathematical Surveys and
  Monographs.\hskip 1em plus 0.5em minus 0.4em\relax Providence, RI: Amer.
  Math. Soc., 1995, vol.~41.

\bibitem{sontag_near_2007}
E.~D. Sontag, ``Monotone and near-monotone biochemical networks,''
  \emph{Systems and Synthetic Biology}, vol.~1, pp. 59--87, 2007.

\bibitem{Tuller2010}
T.~Tuller, A.~Carmi, K.~Vestsigian, S.~Navon, Y.~Dorfan, J.~Zaborske, T.~Pan,
  O.~Dahan, I.~Furman, and Y.~Pilpel, ``An evolutionarily conserved mechanism
  for controlling the efficiency of protein translation,'' \emph{Cell}, vol.
  141, no.~2, pp. 344--54, 2010.

\bibitem{Tuller2007}
T.~Tuller, M.~Kupiec, and E.~Ruppin, ``Determinants of protein abundance and
  translation efficiency in s. cerevisiae.'' \emph{PLOS Computational Biology},
  vol.~3, pp. 2510--2519, 2007.

\bibitem{TullerGB2011}
T.~Tuller, I.~Veksler, N.~Gazit, M.~Kupiec, E.~Ruppin, and M.~Ziv, ``Composite
  effects of gene determinants on the translation speed and density of
  ribosomes,'' \emph{Genome Biol.}, vol.~12, no.~11, p. R110, 2011.

\bibitem{Tuller2010b}
T.~Tuller, Y.~Y. Waldman, M.~Kupiec, and E.~Ruppin, ``Translation efficiency is
  determined by both codon bias and folding energy,'' \emph{Proceedings of the
  National Academy of Sciences}, vol. 107, no.~8, pp. 3645--50, 2010.

\bibitem{wall_contin_frac}
H.~S. Wall, \emph{Analytic Theory of Continued Fractions}.\hskip 1em plus 0.5em
  minus 0.4em\relax Bronx, NY: Chelsea Publishing Company, 1973.

\bibitem{Welch2009}
M.~Welch, S.~Govindarajan, J.~Ness, A.~Villalobos, A.~Gurney, J.~Minshull, and
  G.~C., ``Design parameters to control synthetic gene expression in
  {Escherichia} coli,'' \emph{PLoS ONE}, vol.~4, no.~9, p. e7002, 2009.

\bibitem{infi_HRFM}
Y.~Zarai, M.~Margaliot, and T.~Tuller, ``Explicit expression for the
  steady-state translation rate in the infinite-dimensional homogeneous
  ribosome flow model,'' \emph{IEEE/ACM Trans. Computational Biology and
  Bioinformatics}, vol.~10, no.~5, pp. 1322--1328, 2013.

\bibitem{HRFM_concave}
\BIBentryALTinterwordspacing
Y.~Zarai, M.~Margaliot, and T.~Tuller, ``Maximizing protein translation rate in
  the ribosome flow model: the homogeneous case,'' \emph{IEEE/ACM Trans.
  Computational Biology and Bioinformatics}, 2014, to appear. [Online].
  Available: \url{http://arxiv.org/abs/1407.0207}
\BIBentrySTDinterwordspacing

\bibitem{Zhang1994}
S.~Zhang, E.~Goldman, and G.~Zubay, ``Clustering of low usage codons and
  ribosome movement,'' \emph{J. Theoretical Biology}, vol. 170, pp. 339--54,
  1994.

\bibitem{TASEP_tutorial_2011}
R.~Zia, J.~Dong, and B.~Schmittmann, ``Modeling translation in protein
  synthesis with \mbox{TASEP}: A tutorial and recent developments,'' \emph{J.
  Statistical Physics}, vol. 144, pp. 405--428, 2011.

\bibitem{Zur2013}
H.~Zur and T.~Tuller, ``New universal rules of eukaryotic translation
  initiation fidelity,'' \emph{PLOS Computational Biology}, vol.~9, no.~7, p.
  e1003136, 2013.

\end{thebibliography}

 \end{document}